\documentclass[MSNbibl,nameyear,dvips]{arxstspdf}
\usepackage{flushend}
\usepackage{stfloats}

\volume{26}
\issue{2}
\pubyear{2011}
\firstpage{210}
\lastpage{211}
\doi{10.1214/11-STS338REJ}
\referstodoi{10.1214/10-STS338}

\begin{document}
\begin{frontmatter}

\title{Rejoinder}
\runtitle{Rejoinder}

\begin{aug}
\author{\fnms{Malay} \snm{Ghosh}\corref{}\ead[label=e1]{ghoshm@stat.ufl.edu}}
\runauthor{M. Ghosh}

\affiliation{University of Florida}

\address{Malay Ghosh is Distinguished Professor, University of Florida, 223
Griffin--Floyd Hall, Gainesville, Florida 32611-8545, USA \printead{e1}.}

\end{aug}



\end{frontmatter}

It is a pleasure to receive comments from two very distinguished statisticians
who themselves  have made fundamental contributions to the development of
objective
priors. Their comments clarify many of the ideas presented in this paper,
thereby providing further insight to the selection of objective priors. I~will
respond individually to their comments.

\section*{Bernardo}

I agree with Professor Bernardo that prior elicitation is nearly impossible in
situations which call for very complex models. What I meant to say is that
with years of accumulated data (e.g., for medical practitioners), it is
sometimes possible to elicit a~reasonable prior for certain parameters of
frequent interest (e.g., the cure
probability of a particular drug). In dose-response models, it is often
possible to find meaningful priors for the logistic regression parameters.

I agree with Bernardo that objective Bayesian me\-thods
are unquestionably more appealing than ad hoc frequentist methods. A classic
example is the Behrens--Fisher problem. Also, he is correct in asserting that
even frequentist concepts such as minimaxity,
admissibility, etc. call for Bayesian tools, and objective priors can become
quite handy for such situations. A point to note here, though, is that
since these concepts are
not primarily Bayesian, often the choice leads to quite unappealing priors.
For example, for the binomial proportion, minimaxity demands a $(\sqrt{n}/2,
\sqrt{n}/2$) prior, where $n$ is the sample size. I sincerely doubt that
any practitioner will ever be interested in using such a prior.

I owe an apology to Professor Bernardo for not referring to Berger,
Bernardo and Mendoza. I am also thankful to him for pointing out that in
reference analysis, one does not let the sample size $n$ go to infinity,
but lets $k$, the conceptual number of replicates of the original
experiment, go to infinity.

It was never my intention to advocate priors alternate to Jeffreys in the one
parameter case. My sole objective was to point out that if one considers a
general class of divergence priors, Jeffreys' prior emerges as the solution
in the interior of the parameter space, but not on the boundary. This is
more in the spirit of telling a complete story rather than preaching
something new. For instance, in the binomial case, I do not recommend
necessarily using the $\operatorname{Beta} (1/4,1/4$) prior in preference to Jeffreys'
$\operatorname{Beta}(1/2,1/2)$ prior unless there are other good reasons for using the
former.

I like to point out that in the ratio of normal means problem, the
probability matching criterion does not reproduce the conventional
Fieller--Creasy frequentist solution. This has been exploited in a~very
general framework by Ghosh, Yin and Kim (\citeyear{GhoYinKim03}). Also, I like to add
that while reference priors have general universal appeal, often their
choice is very much dependent on the ordering of parameters. This may be a
daunting task, especially for very complex models. Presumably, one can
salvage such situations by considering prediction rather than estimation.

\section*{Sweeting}

I agree essentially with every single comment made by Professor Sweeting and
indeed thank him for bringing out several important issues barely touched
upon in my article. I take this opportunity to underscore a~couple of the
fundamental arguments that he has put forward.

The first one is the contrast between estimation and prediction.
Bernardo's proper scoring rule is based on the negative of the logarithm
of the prior predictive pdf, geared primarily toward parametric estimation.
In contrast, the negative of the logarithm of the posterior predictive pdf
is ideally suited for prediction. In many situations, it is difficult, if
not impossible to pinpoint the parameter of interest. Predictive inference
for unobserved but potentially observable quantities does not face this
problem, and often is the most desired mode of inference. The currently
popular neural nets and machine learning techniques aim solely toward
prediction. A~more classical example is finite population sampling where
the goal is to find the predictive distribution of the unobserved given
the observed.

The second important point is that often the prior can overshadow the data.
The simple (albeit artificial) example put forward by Professor Sweeting
amply demonstrates this. Implicit in this is the fact that nonjudicious
selection
of priors by nonexperts can lead to meaningless inference far removed from
reality. While modern statistics in general, and Bayesian statistics in
particular, is trying to tackle high-dimensional complex data problems,
the existing methodology may not always be adequate to provide the right
solution. However objectionable to a purist, it may genuinely be necessary
to invent data-dependent priors from a pragmatic standpoint. I may add, though,
that if we subscribe to this last dictum, there is no particular reason to
criticize any particular mode of selection of objective priors as long as
it works.

\section*{Final Remarks}

I strongly believe that we will never be able reach a consensus on the
selection of a default prior which\ will work well under all situations. Also,
I do not see any need for such an agreement. A very prudent thing is to have a
large number of the so-called ``objective'' priors in one's toolkit, and see
which one is most appropriate in a particular instance. Also, it is important
to replenish the toolkit from time to time. I agree with Sweeting that the term
``subjective'' is possibly better than ``objective'' in  spite of the latter's
current global usage. Finally, I thank the discussants for their valuable
comments, and hope that the dialogue will continue in the future.


\end{document}